# Analyzing the Reporting Error of Public Transport Trips in the Danish National Travel Survey Using Smart Card Data

**A Preprint**

**Georges Sfeir**
Department of Technology, Management and Economics
Technical University of Denmark, Kongens Lyngby, Denmark, 2800
Email: geosaf@dtu.dk

**Filipe Rodrigues**
Department of Technology, Management and Economics
Technical University of Denmark, Kongens Lyngby, Denmark, 2800
Email: rodr@dtu.dk

**Maya Abou Zeid**
Department of Civil and Environmental Engineering
American University of Beirut, Beirut, Lebanon, 1107 2020
Email: ma202@aub.edu.lb

**Francisco Camara Pereira**
Department of Technology, Management and Economics
Technical University of Denmark, Kongens Lyngby, Denmark, 2800
Email: camara@dtu.dk

## ABSTRACT

Household travel surveys have been used for decades to collect individuals and households' travel behavior. However, self-reported surveys are subject to recall bias, as respondents might struggle to recall and report their activities accurately. This study examines the time reporting error of public transit users in a nationwide household travel survey by matching, at the individual level, five consecutive years of data from two sources, namely the Danish National Travel Survey (TU) and the Danish Smart Card system (Rejsekort). Survey respondents are matched with travel cards from the Rejsekort data solely based on the respondents' declared spatiotemporal travel behavior. Approximately, 70% of the respondents were successfully matched with Rejsekort travel cards. The findings reveal a median time reporting error of 11.34 minutes, with an Interquartile Range of 28.14 minutes. Furthermore, a statistical analysis was performed to explore the relationships between the survey respondents' reporting error and their socio-economic and demographic characteristics. The results indicate that females and respondents with a fixed schedule are in general more accurate than males and respondents with a flexible schedule in reporting their times of travel. Moreover, trips reported during weekdays or via the internet displayed higher accuracies compared to trips reported during weekends and holidays or via telephone interviews. This disaggregated analysis provides valuable insights that could help in improving the design and analysis of travel surveys, as well accounting for reporting errors/biases in travel survey-based applications. Furthermore, it offers valuable insights underlying the psychology of travel recall by survey respondents.

# 1 INTRODUCTION

For decades, researchers and transport planners have relied on household travel surveys (HTS) or travel diaries to collect data on the travel behavior of individuals and households. Advancements in technology have brought a revolution in the domain of data collection. Particularly, passively-generated big data such as GPS, mobile phone , and smart card data, provide opportunities to complement and enrich traditional surveys (Bonnel and Munizaga 2018; Callegaro and Yang 2017; Deschaintres et al. 2022; Miller et al. 2018) as every data collection technique (traditional or emerging) has advantages and drawbacks.

Traditional household travel surveys provide comprehensive contextual information about individuals' travel behavior, travel preferences, and socio-economic characteristics and additionally provide the opportunity to collect attitudinal variables that are not usually available in other data sources (Bayart et al. 2009). Traditional household travel surveys are also flexible as they can be tailored by researchers to specific research objectives. However, these surveys rely on the respondents' ability to accurately report details about their activities and trips such as number of trips, departure time, origin and destination, etc. Unfortunately, people have a well-known tendency to inaccurately report such information (Stopher et al. 2007). Researchers and transport planners have long been aware of the recall bias in self-reported surveys caused by the participants' inability to recall and report their travel activities accurately (Clarke et al. 1981). However, a relatively fair assessment of the actual margin/level of error in self-reported travel surveys has only become possible with the recent implementation of GPS surveys and smart card systems (Stopher et al. 2007).

Both GPS-based surveys and smart card systems automatically collect precise and real-time data on travel behavior, minimizing the reliance on individuals' memory and mitigating, although to different levels, recall bias. Nonetheless, the two systems are fundamentally different. Although smart card systems provide only partial data on public transport passengers' travel behavior/patterns (e.g., trip origin and destination are unknown), they still offer several advantages over GPS and mobile phone surveys. First, smart card data is not limited by equipment or battery life, allowing for longer data collection periods (Spurr et al. 2015). Second, the high penetration and usage rate of smart card systems in many cities and countries allows them to cover almost the entire population of travelers. In contrast, GPS surveys may suffer from a sample selection bias, as participants who agree to participate in the survey and carry a GPS device or install a smartphone app for collecting travel diaries during their travel might have different socio-economic characteristics and travel patterns compared to those who decline to participate (Bricka and Bhat 2006; Lugtig et al., 2022). Finally, no additional effort is required from travelers other than validating the fare by tapping in and, in some cases, tapping out compared to GPS surveys where participants are required to self-verify their trips (Li and Shalaby 2008; Zhao et al. 2015), something that might introduce self-reporting errors in the data.

## 1.1 Smart Card Data and Travel Surveys

Several studies have recently tried to compare and/or integrate household travel surveys and smart card data at the population/aggregate level to assess the advantages and disadvantages of both data collection methods and/or attempt to alleviate reporting errors in travel surveys (Su et al. 2022). For instance, many studies investigated the potential of travel underreporting in travel surveys. Ridership of the Montréal, Canada, subway system was investigated by comparing an average weekday of travel demand data from the 2008 Montréal household travel survey and one day of smart card transactions from 2010 (Spurr et al. 2014). Results showed that the survey accurately represents daily subway ridership but overestimates subway boardings during peak hours by 24%. To correct this overestimation, the weights of home-based trips were calibrated by the actual entry volumes at subway stations during peak periods. However, even after the recalibration, it was found that the survey underestimates off-peak demand by 21% likely due to nonresponse of specific groups and underreporting of non-home-based trips by respondents. Spurr et al. (Spurr et al. 2018) also compared two datasets describing the public transit demand in the Greater Montréal region, the Montréal household travel survey and smart card transactions. Household travel survey data from the fall of 2013 was extracted to construct an average weekday data while smart card data was collected on a specific day from October 2013. The authors compared the structure of transit travel demand (e.g., temporal distribution of trips during a typical weekday, types of fare product used, spatial and temporal distribution of trips over the multiple networks serving the metropolitan area) in both datasets and found that the household travel survey over-represents symmetrical travel patterns observed during peak periods and between the suburbs and downtown while neglecting other travel patterns. Other studies also compared the Montréal household travel survey data with smart card data at the aggregate level. Trépanier et al. (2009) found that the 5% sampling rate in the household survey was insufficient for capturing significant daily temporal variations and ridership of specific bus lines. Chapleau et al. (2018) showed that non-home-based trips and trips made for short duration activities that are mainly conducted during off-peak periods are under-reported in the household travel survey. Furthermore, the findings indicated that the overestimation of public transit trips during peak periods can be attributed to the use of weighting factors to address the underrepresentation of the 20-29 age demographic group of the population, which is difficult to reach in a telephone-based household travel survey. Deschaintres et al. (2022) applied a weighting method to expand the representativeness of the OD household travel survey of Montréal from a typical daily trip dairies to a four-month period. By comparing the results to smart card and count data, the authors showed similar linear trends and weekly variations in the daily use of cars and subways. Public transit OD trip matrices were also derived for Lyon, France, from a household travel survey, a large-scale OD survey, and an entry-only smart card data (Egu and Bonnel 2020). Results showed that although the three matrices share some similarities, they have significant differences that must be acknowledged and investigated. For instance, the household travel survey tends to underestimate public transport trips by approximately 30% while overestimating long-distance and multi-leg trips during peak hours.

While most of the previous studies have compared travel surveys and smart card data at the population/aggregate level, very few efforts have been made to match and compare the two data collection methods at the individual level (Riegel and Attanucci 2014; Spurr et al. 2015; Su et al. 2022). Riegel and Attanucci (2014) compared smart card transactions with London Travel Demand Survey (LTDS) responses of individuals who willingly provided their smart card numbers to enable their identification in the smart card data. Only around half of the reported trip legs over a 9-month period were matched to smart card transactions. In addition, large differences in duration and start time were noticed with an average start time difference of more than an hour. Spurr et al. (2015) applied a methodology based on spatiotemporal filters to match smart card data with household travel survey responses of individuals who were not asked to provide their smart card numbers. The authors were able to match roughly 50% of HTS transit users and identify as such three different categories of survey respondents: those who report almost accurately their travel, those who underreport their travel, and those who report typical trips instead of actual ones. However, the study is limited to one day of travel diaries from the 2013 Montréal household travel survey and as such the derived typology is not exhaustive. Su et al. (2022) compared self-reported travel data from the MIT commuting survey with smart card transactions and on-campus parking records of MIT employees both at the aggregated and individual level. Results showed some level of inconsistency between the datasets and that the overreporting and underreporting of commuting patterns are associated with certain individual characteristics such as age and employment type. However, the study is limited to a particular category of the population (MIT employees). In addition, the two datasets were not matched on a daily basis nor daily discrepancies were assessed. The above-mentioned studies are summerazed in Table 1.

**Table 1: Summary of studies comparing HTS and SC data**

| Study | Methodology | Key Findings |
|---|---|---|
| Spurr et al. (2014) | Aggregate comparison of HTS and SC data | Over-reporting of peak-hour travel by 24%, under-reporting of off-peak demand by 21% |
| Spurr et al. (2018) | Aggregate comparison of HTS and SC data | Over-reporting of symmetrical travel patterns during peak periods, between the suburbs and downtown |
| Trépanier et al. (2009) | Aggregate comparison of HTS and SC data | Insufficient sampling rate in HTS for capturing daily temporal variations |
| Chapleau et al. (2018) | Aggregate comparison of HTS and SC data | Under-reporting of non-home-based trips and short-duration activities, bias from weighting factors |
| Deschaintres et al. (2022) | Aggregate comparison of HTS and SC data | Weighted HTS showed similar trends and weekly variations as in SC data |
| Egu and Bonnel (2000) | Aggregate comparison of HTS and SC data | Under-reporting of PT trips |
| Riegel and Attanucci (2014) | Disaggregate/Individual comparison of HTS and SC data | Only 50% match rate between reported trips and SC data, significant timing discrepancies (average start time difference of more than 60 minutes) |
| Spurr et al. (2015) | Disaggregate/Individual comparison of HTS and SC data | One day of data. Around 50% match rate. Three categories of respondents were identified with different reporting accuracy levels |
| Su et al. (2022) | Disaggregate/Individual comparison of HTS and SC data | Inconsistencies in reporting commuting patterns, correlation with age and employment type |

In summary, previous studies have mostly compared traditional survey data and smart card transactions at the aggregate level, while those few who delved into a more detailed comparison at the disaggregate level were often limited by time constraints such as one day of data, a few months, or data from different time periods for each dataset. Consequently, there is a lack of exhaustive and comprehensive comparison between the two different data collection methods at the individual level, specifically in quantifying reporting errors in travel surveys and their correlation with socio-economic and demographic characteristics. This paper aims to fill this gap as it tries to match 5 consecutive years (2018 to 2022) of smart card data and household travel survey for the entirety of Denmark with the objective of quantifying the reporting error of public transport users in the Danish national travel survey and investigating the relationships between these errors and various socio-economic characteristics of travelers. Such analysis should yield valuable insights that can help in improving the design and analysis of travel surveys, leading to more accurate data collection techniques. It would also help researchers in accounting for reporting errors/biases when using travel survey data. Furthermore, it would offer valuable insights underlying the psychology of travel recall by survey respondents. To the best of our knowledge, this is the largest and most in-depth analysis of this sort thanks to the sheer size of the datasets used and the unique characteristics of the Danish national travel survey.

The remainder of this paper is organized as follows. First, the smart card data and travel survey are presented. Second, the matching process between the two datasets is presented. Next, results of the matching process and a statistical analysis are presented and discussed. Finally, the findings and future extensions of this work are discussed.

## 2 DATA

This section describes the two datasets used in this study, the smart card Rejsekort data and the Danish National Travel Survey (TU Data).

### 2.1 Smart Card - Rejsekort - Data

The Danish Rejsekort (travel card in English) is the nationwide smart card system for traveling by public transport in Denmark. Under this system, passengers must tap-in at their origins and transfer locations and tap-out at their destinations. The Rejsekort system covers all public transport modes (buses, metros, and trains), transport operators, and travel zones in Denmark (Rejsekort, 2023). Each Rejsekort transaction stores information on the type of transaction (tap-in, transfer, or tap-out), time and location of the transaction, type of the card, and fake card ID (Rejsekort IDs are pseudo-anonymized for privacy concerns). For this study, the whole Rejsekort data for all Denmark from 2018 to 2022 are used. Table 2 and Figure 1 show the number of cards with one to four or more trips per average day[1] from 2018 to 2022. Although the number of smart card trips is considered large compared to number of reported trips in the TU data (refer to section 2.2), the average number of trips per origin-destination (OD) per day in 2018 varies between 2.08 and 2.78 trips/OD with a median of 1 trip/OD/day and a standard deviation of 4.59 to 10.06 trips/OD/day (Figure 2). Similar values are found from 2019 to 2022. More descriptive statistics are shown in Appendix A.

---

[1] All days on which respondents reported PT trips in the TU survey were included in the calculation of the average day.

**Table 2: Number of cards with 1 to 4+ trips per average day from 2018 to 2022**

| Year | Cards with 1 PT trip/day | Cards with 2 PT trips/day | Cards with 3 PT trips/day | Cards with 4+ PT trips/day | Total |
|---|---|---|---|---|---|
| 2018 | 107,590 | 126,398 | 17,780 | 6,181 | 257,948 |
| 2019 | 90,027 | 109,512 | 15,427 | 5,597 | 220,563 |
| 2020 | 101,731 | 124,636 | 16,669 | 5,906 | 248,942 |
| 2021 | 97,798 | 125,197 | 15,610 | 5,423 | 244,027 |
| 2022 | 118,856 | 151,091 | 21,010 | 7,783 | 298,740 |

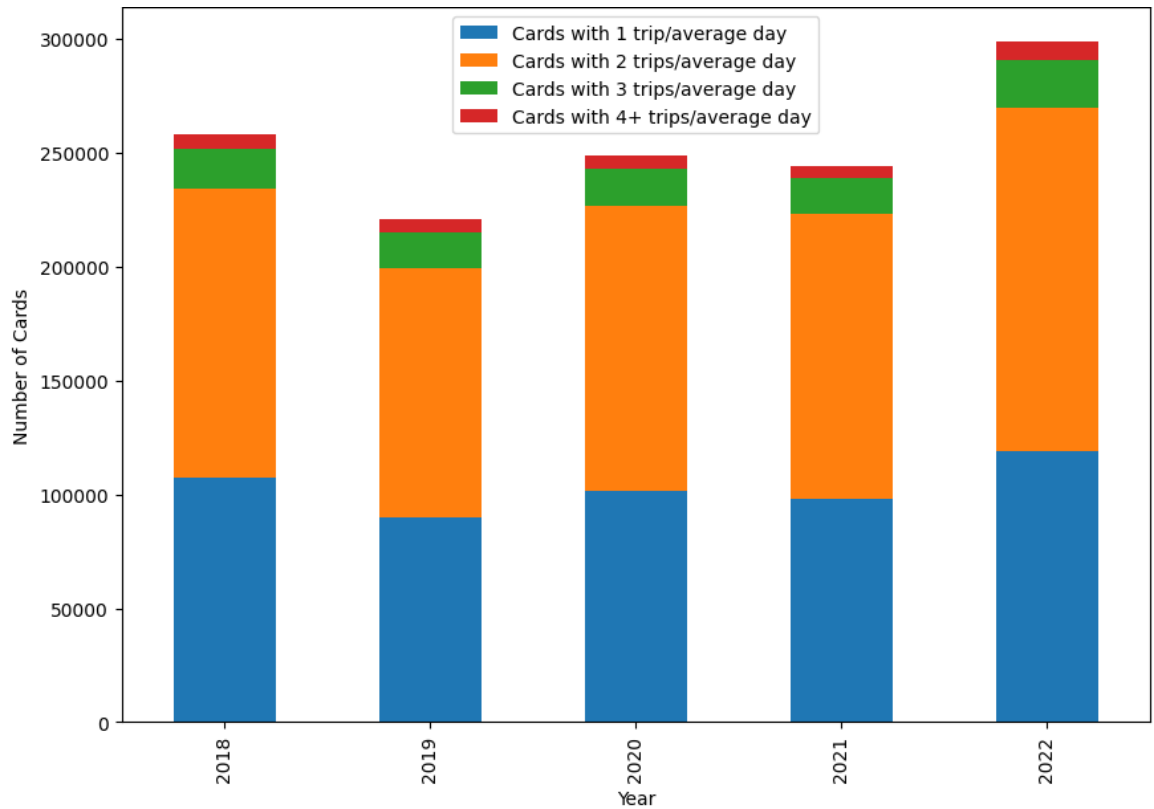

**Figure 1: Number of cards with 1 to 4+ trips per average day from 2018 to 2022**

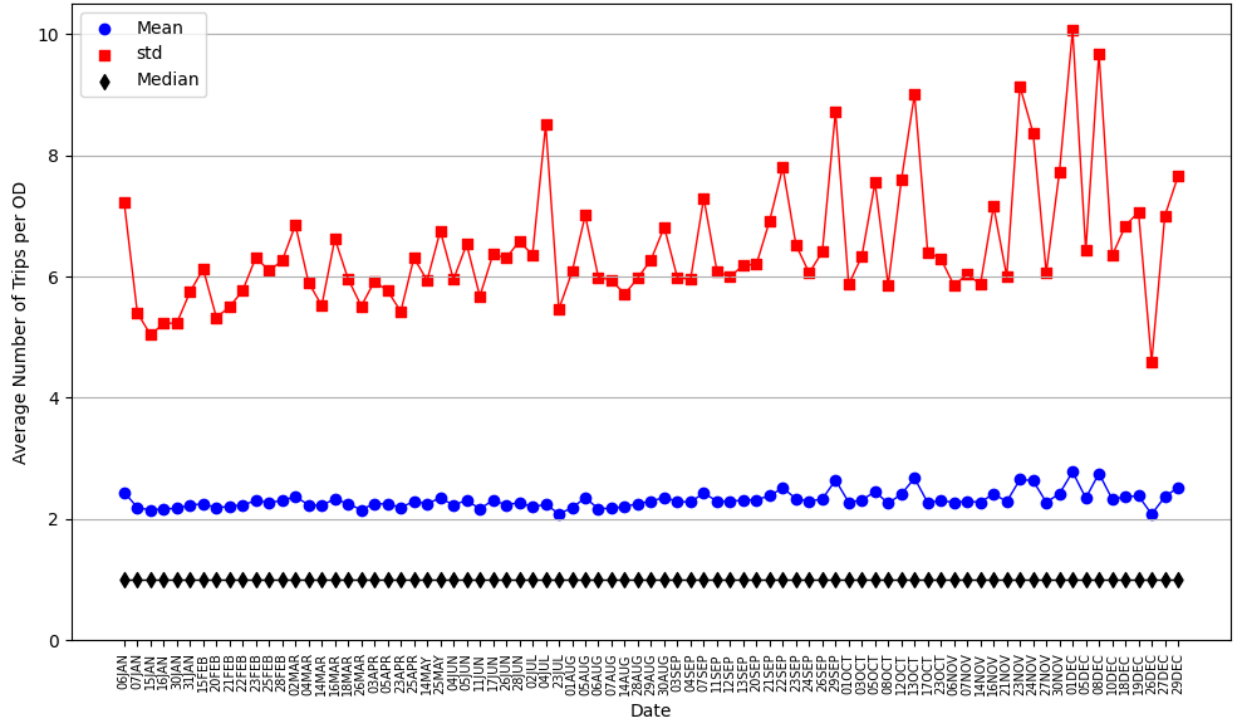

**Figure 2: Number of trips per OD per day across all days on which TU respondents reported PT trips in 2018**

## 2.2 Danish National Travel Survey - TU Data

The Danish National Travel Survey (or Transportvaneundersøgelsen, TU) is an annual survey that aims to capture travel diaries of a representative sample of the Danish population aged 6 years and above (Christiansen and Baescu 2023). The survey is conducted on random days throughout the year and is either answered via telephone (80%) or internet (20%). Participants are asked to provide detailed descriptions of all their trips undertaken using both private and public modes of transportation on the day prior to the interview (e.g., selected modes, departure time, trip duration, distance traveled, and trip purpose) in addition to their own and households' socio-economic characteristics (e.g., gender, age, income, location of residence and workplace). For this study, public transport trips conducted with a Rejsekort card between 2018 and 2022 are selected for matching with trips from the Rejsekort data during the same period. The collected information on public transport trips is sufficiently detailed to enable matching the reported trips with the "actual" trips recorded in the Rejsekort data. For each public transport trip in the TU survey, respondents are asked to provide start and end times, all modes (train, bus, metro) used for all legs of the trip, waiting time, bus line, length and travel time for each mode and leg, names of boarding, transfer, and alighting metro and train stations, etc. Table 3 summarizes the number of public transport users and trips reported as conducted by a Rejsekort card in the TU data between 2018 and 2022. We only try to match TU respondents who reported two or three trips per day with corresponding cards from the Rejsekort data. This is based on the understanding that the likelihood of finding multiple individuals/cards with exactly the same two or three trips per day (same tap-in and tap-out locations and times) is expected to be very low. On the other hand, for TU respondents who reported only one trip per day, there is a higher probability of identifying multiple matches in the Rejsekort data, where cards have transactions with the same tap-in and tap-out locations and times. Matching these cases would require additional information. As for respondents who reported four to eight trips per day, they account for approximately 2% of the sample. Matching such a small subset is computationally expensive and highly unsuccessful. This is due to the tendency of these respondents to report shorter trips nested within longer trips, where they may not have tapped-in and/or tapped-out consistently, making successful and reliable matching challenging. In total, 3,750 public transport trips were reported by 2,116 respondents as made using a Rejsekort card between 2018 and 2022. Out of those, 1,208 respondents reported two trips per day while only 128 respondents reported three trips per day. Therefore, the total number of respondents used for matching is 1,336 which corresponds to 2,800 trips (Table 3).

**Table 3: Public transport trips and Rejsekort users in TU data from 2018 to 2022**

| Year | Reported PT Trips | Respondents with PT Trips | Respondents with 1 PT trip/day | Respondents with 2 PT trips/day | Respondents with 3 PT trips/day | Respondents with 4+ PT trips/day |
|---|---|---|---|---|---|---|
| 2018 | 732 | 427 | 169 | 223 | 24 | 11 |
| 2019 | 786 | 442 | 150 | 255 | 28 | 9 |
| 2020 | 681 | 388 | 141 | 216 | 19 | 12 |
| 2021 | 664 | 374 | 125 | 218 | 24 | 7 |
| 2022 | 887 | 485 | 144 | 296 | 33 | 12 |
| Total | 3,750 | 2,116 | 729 | 1,208 | 128 | 51 |

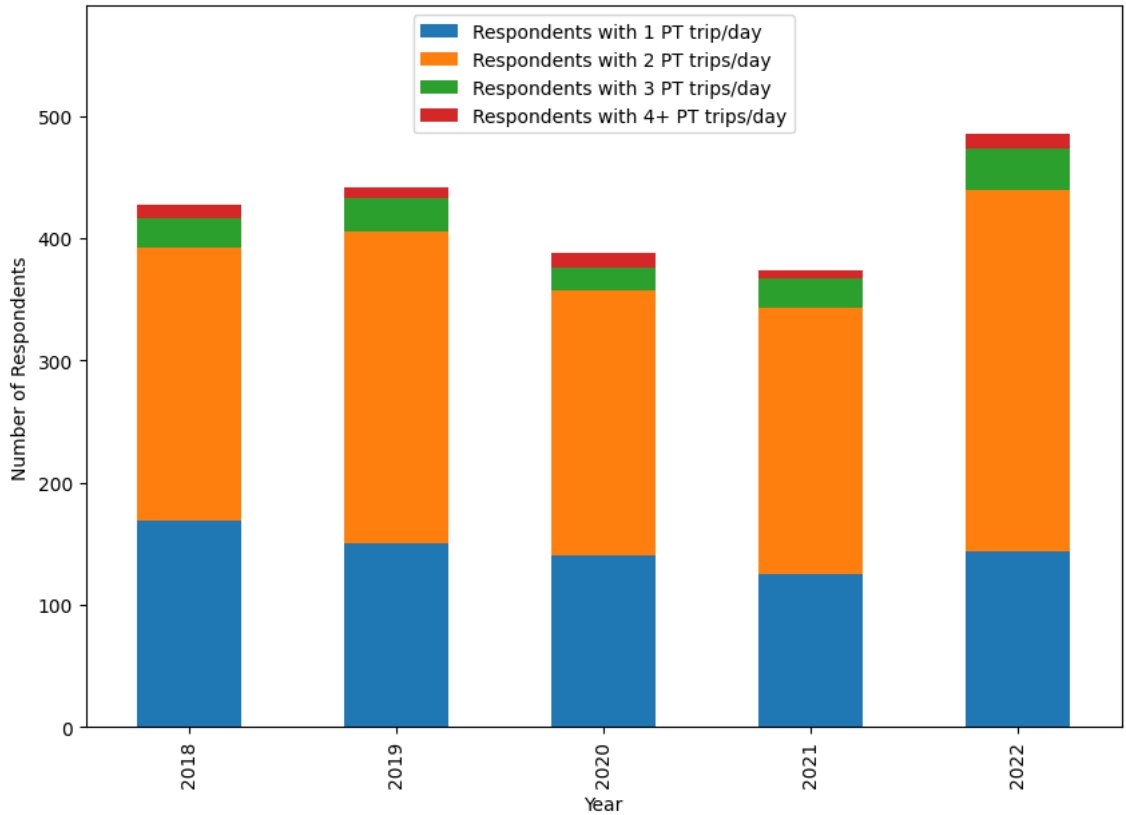

**Figure 3: Respondents by number of reported PT trips with Rejsekort cards from 2018 to 2022**

## 3 MATCHING

For the purpose of matching reported TU trips with Rejsekort trips, we classify the respondents from the TU data into three categories - train, bus, and mixed users - and perform the matching for each category separately. The train users are respondents who reported 2 or 3 trips where all legs of all trips were made by train or metro. The bus users are respondents who reported 2 or 3 trips where all legs of all trips were made by bus. Mixed users are respondents who reported 2 or 3 trips of mixed bus and train legs. This category includes users who made at least one trip completely by bus and one completely by train or metro, as well as users who made at least one trip with mixed bus and train or metro legs.

In order to develop a matching process between the two datasets, it is necessary to understand the definition and characteristics of a trip category in each dataset. Figure 4 shows the definition and attributes of a train trip as described in both datasets. Actual origin and destination are not known in the TU dataset for privacy concerns. However, all time instants (e.g., departure time, waiting time, arrival time, etc.) of a train trip from origin to destination are known as well as the names of the boarding, transfer, and alighting stations. As for the Rejsekort data, origin and departure times are not known and neither are the destination and arrival times. Instead, only names and times, precise to the nearest second, of tap-in and tap-out stations are known. A traveler taking a train can tap-in any time between arrival at the station and right before boarding the train. However, it is assumed that travelers usually tap-in when arriving at the station. Therefore, we define the reported arrival time to the first station in the TU data as the possible tap-in time. As for tap-out time, it corresponds to alighting time of the last train as travelers usually tap-out right after leaving the train.

Given that TU respondents are not asked to provide their Rejsekort IDs and the IDs in the Rejsekort dataset are themselves pseudo-anonymized, matching respondents in the TU data to smart cards from the Rejsekort data can only be performed based on their observed travel behaviors. The matching methodology for the train category is described as follows:

1. Given an individual $n$ who reported in the TU survey $I_n:\{2,3\}$ trips by train during a specific day $d$, get the names of the first (boarding) and last (alighting) stations of each trip $i \in I_n$ from TU data

2. Find $I_n$ trips in the Rejsekort data made by the same Rejsekort card that match the names of the stations from the TU data during day $d$

3. Compute absolute time difference, $\Delta T_n$, between Tap-in/Tap-out times from Rejsekort data and Arrival to first/last stations from TU data of all trips as follows:

$$\Delta T_n = \sum_{i=1}^{I_n} \Delta T_{n,First\_St_i} + \sum_{i=1}^{I_n} \Delta T_{n,Last\_St_i}$$

where $\Delta T_{n,First\_St_i}$ is the absolute time difference of trip $i$ between arrival time to first station from TU data and tap-in time at first station from Rejsekort data; $\Delta T_{n,Last\_St_i}$ is the absolute time difference of trip $i$ between alighting time at last station from TU data and tap-out time at last station from the Rejsekort data.

4. If there is more than one match, select the one with the smallest $\Delta T_n$

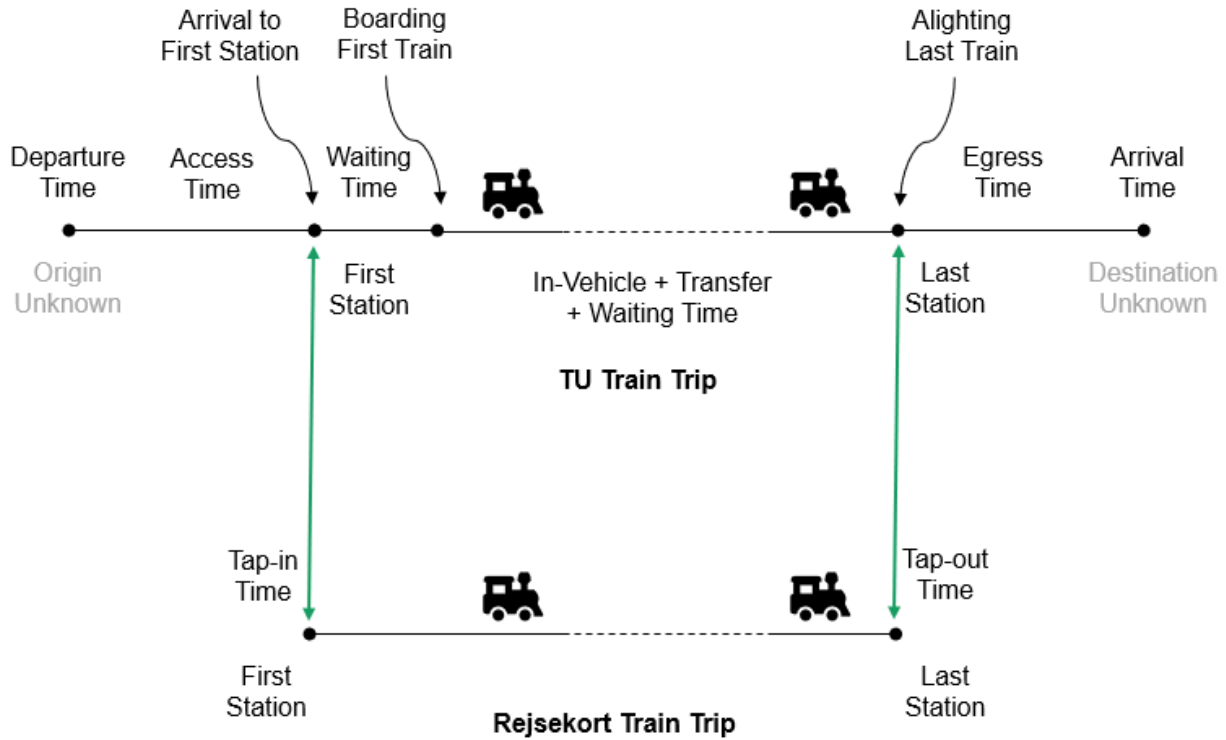

**Figure 4: Train trip attributes as recorded by TU and Rejsekort data**

Similarly to a train trip, the actual origin and destination of a bus trip are not observed in the TU dataset for privacy concerns. In addition, names of boarding and alighting bus stops are not recorded. Instead, respondents are asked about the boarding and alighting bus lines. As for the Rejsekort data, both bus lines and bus stops are recorded (Figure 5). Therefore, the matching process for the bus category is based on the bus lines instead of bus stops and follows the same steps previously mentioned for the train trips as follows: 1) get the names of bus lines at the first (boarding) and last (alighting) stops instead of names of bus stops; 2) find trips in the Rejsekort data made by the same Rejsekort card that match the names of the bus lines; 3) compute the absolute time difference $\Delta T_n$ according to equation 1 but where $\Delta T_{n,First\_St_i}$ / $\Delta T_{n,Last\_St_i}$ are the absolute time difference between boarding time of first bus / alighting time of last bus from TU data and tap-in time at first stop / tap-out time at last stop from Rejsekort data; 4) if there is more than one match, select the one with the smallest $\Delta T_n$.

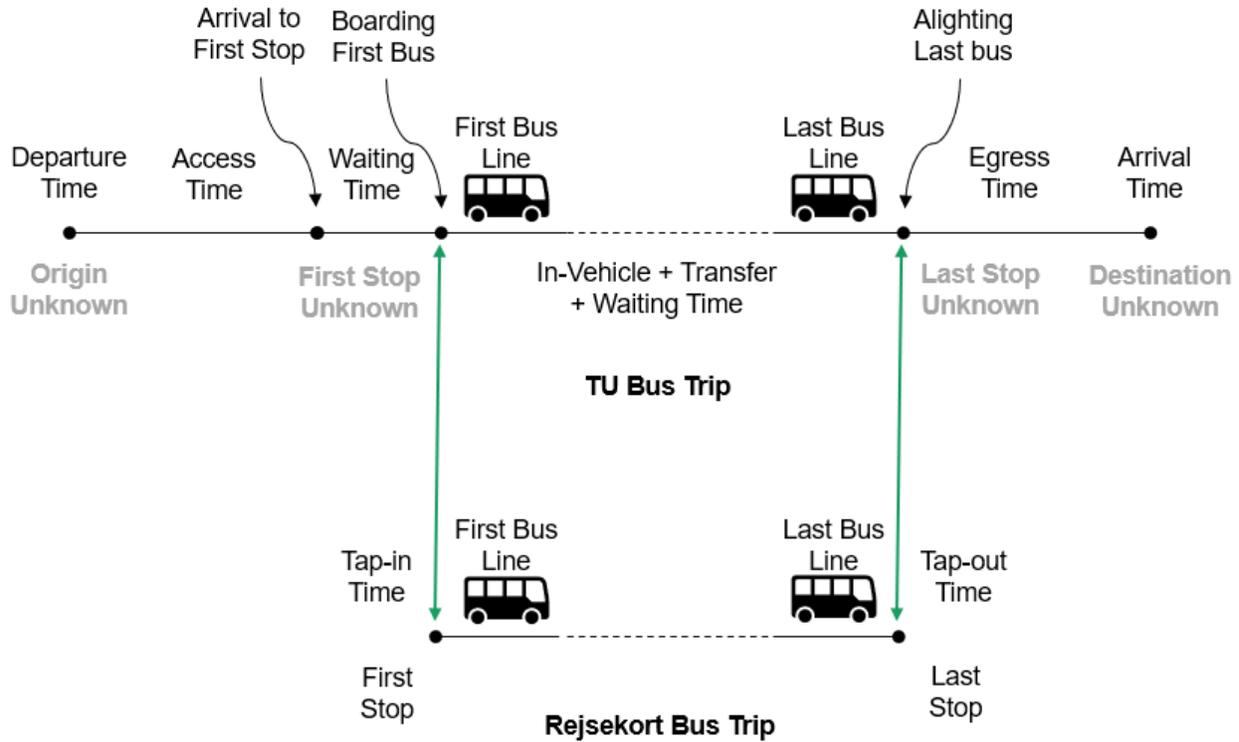

**Figure 5: Bus trip attributes as recorded by TU and Rejsekort data**

Finally, the matching of the mix category is based on a combination of the train (stations) and bus (lines) matching processes.

## 4 RESULTS & STATISTICAL ANALYSIS

This section presents the results of the matching process between the two datasets in addition to a statistical analysis.

### 4.1 Matched Sample

Using the matching process described in the previous section, 70.51% of TU respondents with 2 or 3 PT trips were successfully matched with at least one Rejsekort card with equivalent sequences of tap-ins/tap-outs (Table 4). The matched respondents can be divided into the following categories: 507 respondents (53.82%) with train trips, 304 (32.27%) with bus trips, and 131 (13.91%) with mixed trips. In addition, 898 matched respondents (95.33%) reported 2 trips while only 44 matched respondents (4.67%) reported 3 trips. Therefore, the total number of matched trips is 1,928. The relatively high matching rate (70.51%) compared to roughly 50% from (Spurr et al. 2015) could be attributed to the different fare validation requirements upon exiting the public transport systems in Montréal and Denmark. In Montréal, fare validation is not required upon leaving the system and as such alighting locations are not available in the Montréal smart card data and are instead imputed through the analysis of transaction chains. On the other hand, the Rejsekort system in Denmark requires travelers to tap-out at their alighting stations/stops, which provides accurate information about the alighting locations in the Rejsekort data. Finally, unmatched trips in this study can be attributed to several potential factors such as people reporting incorrect train/metro stations or bus lines, potential data entry mistakes by interviewers, people forgetting to tap-out at the alighting station/stop, etc.

**Table 4: Matching results**

| Year | TU Respondents with 2 or 3 PT trips | Matched TU respondents with specific Rejsekort cards | Matching % |
| --- | --- | --- | --- |
| 2018 | 247 | 169 | 68.42% |
| 2019 | 283 | 217 | 76.68% |
| 2020 | 235 | 165 | 70.21% |
| 2021 | 242 | 176 | 72.73% |
| 2022 | 329 | 215 | 65.35% |
| Total | 1,336 | 942 | 70.51% |

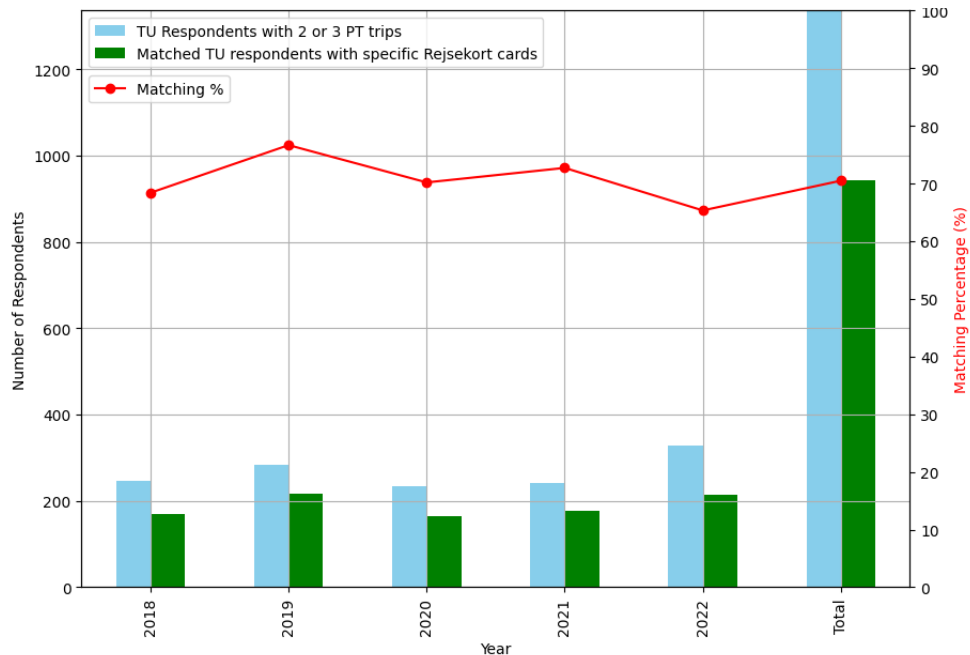

**Figure 6: Matching results**

A Shapiro test was applied to several time difference variables ($ΔT_n$, $average\ ΔT_n$, $ΔT_{n,First\_St}$) to assess the normality of the data. None of the variables are normally distributed at a 99% level of confidence ($p < 0.01$), indicating that non-parametric tests should be used for statistical analysis. The dependent variable of interest for the statistical analysis is the absolute time difference at the first stop/station $ΔT_{n,First\_St}$. We also call this dependent variable the time reporting error of respondents in the TU survey or start time difference.

The reporting error of the matched trips varies between 0 and around 8 hours with a mean of 33.87 minutes and a standard deviation of 65.12 minutes. While these values are relatively high, they are still 50% lower than those from (Riegel and Attanucci 2014) who reported a mean of 61.2 minutes, a standard deviation of 151.7 minutes and start time differences ranging from 0 to 15 hours. It is highly unlikely that a respondent misreported their departure time by 8 hours. Therefore, to lessen the impact of outliers and given that the data is not normally distributed but instead is positively skewed (Figure 7), relying on the median and quartiles would provide a more accurate representation of the data. The reporting error has a median of 11.34 minutes and an Interquartile Range (IQR[2]) of 28.14 minutes. It should be noted that in most stations in Denmark, the time needed to move from the entry to the platforms typically takes less than a minute. Moreover, the temporal resolution in the TU survey is 5 minutes[3]. A median reporting error of 11.34 minutes is more than twice the temporal resolution and as such cannot be ignored or solely attributed to the 5-minute discretization of time in the TU survey. Several studies have documented that most respondents tend to round their departure and arrival times to multiples of 5, 15, and 30 minutes (Rietveld 2002; Stopher et al. 2007). Such large rounding scales could introduce biases into any analysis based on national travel surveys, particularly when probabilities of rounding upward and downward do not balance out (Varela et al. 2018).

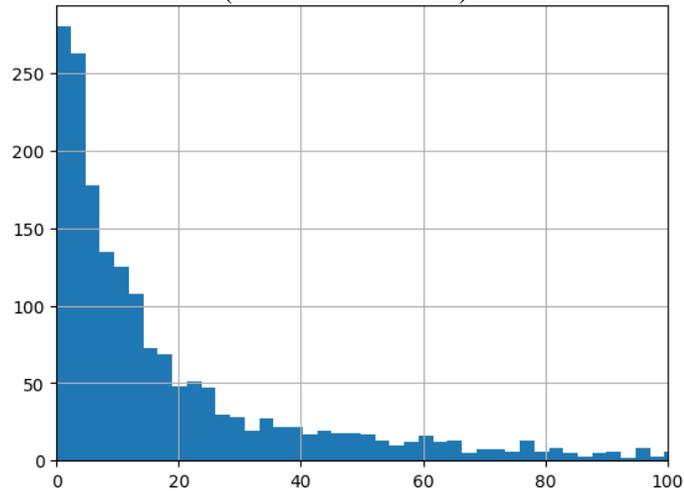

**Figure 7: Distribution of the absolute time difference at the first stop/station $ΔT_{n,First\_St}$ (in minutes)**

---

[2] IQR is the difference between the third (75th percentile) and first (25th percentile) quartiles ($Q_3 – Q_1$). It measures the spread of the middle 50% of the data. It is robust against outliers and an alternative to the standard deviation in case of extreme values.

[3] When selecting time in the TU survey, respondents can choose from a list of 5-minute bins (e.g., 9:05, 9:10 etc.)

## 4.2 Statistical Analysis

Several non-parametric tests were used to compare the dependent variable of interest against different socio-economic variables and their levels. First, the Mann-Whitney U-Test, a non-parametric statistical alternative to the two-sample t-test, also known as the Wilcoxon Rank Sum test, was used to compare two independent groups (e.g., a socio-economic variable with two categories). Second, the Kruskal-Wallis H test, a non-parametric statistical alternative to the one-way ANOVA, was used to compare the distribution of more than two independent samples (e.g., a socio-economic variable with three categories). Finally, the Wilcoxon Signed Rank test, a non-parametric alternative to the paired samples t-test, was used to compare two dependent samples.

*4.2.1 Two-level variables*

In this section, differences between different groups of the population were investigated using the Mann-Whitney U-Test. Descriptive statistics for each group in addition to the p-values of the Mann-Whitney U-test are presented in Table 5. Results show that there are statistical differences at either the 99% or 95% level of confidence between the categories of all the variables. There are statistical differences between males and females at the 99% level of confidence with a higher median value for males (13.33 minutes) compared to females (10.17 minutes), thus indicating that females are in general more accurate than males in recalling and reporting their time of travel. This is in line with insights from the field of psychology that women usually perform better than men on episodic memory[4] tasks (Herlitz et al. 1997). A statistically significant difference in the medians of the two categories of Day Type 1 is found at the 95% level of confidence, suggesting that people are more accurate in reporting their time of travel during weekdays (11 minutes) than during weekends (13.23 minutes). More specifically, people are more accurate in reporting their time of travel during normal weekdays than during weekends and weekdays with holidays (Day Type 2). This is intuitive and expected as people usually follow a predictable routine during the week (e.g., going to work or school), which makes it easier for them to recall and report their time of travel accurately. In contrast, weekends and holidays usually involve a less structured routine with a wider range of nonrepetitive activities (e.g., leisure and social activities), which can make it more challenging for respondents to remember and accurately report their time of travel. Moreover, we categorize respondents according to their schedule flexibility into two groups: those with fixed schedules (e.g., students, employees) and those with flexible schedules (e.g., unemployed, pensioner). A statistically significant difference between the medians of the two groups is evident at the 99% level of confidence. Specifically, respondents with flexible schedules (14.59 minutes) are less accurate in reporting their time of travel in comparison to respondents with fixed schedules (10.78 minutes). This further supports the previous finding that people on average are more accurate during weekdays than during weekends and holidays, which tend to involve greater schedule flexibility. It is to be noted that approximately 63% of trips made by respondents with fixed schedule can be categorized as "must" travel (e.g., work trips, education trips, visit to doctors etc.). In contrast, around 62% of trips conducted by respondents with flexible schedule can be considered as "lust/leisure" trips (e.g., shopping, leisure, sports etc.). This might be attributed to the fact that unemployed individuals and pensioners tend to have more free time, leading to a higher frequency of leisure trips. A statistically significant difference at the 99% level of confidence is also observed for interview type, with internet-based responses having a much lower median (7.23 minutes) than telephone-based responses (12.72 minutes). Internet-based surveys usually incorporate visual aids (e.g., maps or timelines), which assist respondents in remembering and reporting their time of travel. Furthermore, internet-based surveys provide respondents with the flexibility to answer at their own pace and convenience, allowing them more time to recall and report their answers without feeling rushed or distracted/interrupted as they might in telephone-based surveys. Moreover, the elimination of interviewers in internet-based surveys has been shown to minimize social desirability bias (Braunsberger et al. 2007; Couper 2000) and thus lead to higher reporting accuracy. Finally, trips are divided into two categories, Jutland and Zealand/Funen, according to the location of their origins and destinations. A statistically significant difference at the 99% level of confidence is evident (p-value = 0.000) between the median values of the two geographical locations with trips conducted in Jutland having a much lower median value (7.63 minutes) than those conducted in Zealand and Funen (12.31 minutes). Note that Zealand is the most populous island in Denmark and includes the capital Copenhagen. Zealand, and to some extent Funen, are characterized by urban and fast-paced environments while Jutland is predominantly rural. As such, Jutland experiences, in general, a lower bus frequency compared to Jutland, while metro is only available in Copenhagen. This could explain the finding that people in Jutland are more accurate than those from Zealand and Funen as with lower bus frequencies travelers might be more aware of the bus schedules to minimize waiting times. Note that buses and trains are schedule-based while metros operate on a headway-based system.

---

[4] Episodic memory is the ability to recall past events or experiences at particular times and spaces.

A sensitivity analysis was also performed to test whether the Mann-Whitney U-Test conclusions hold if trips with high reporting errors are excluded. The statistical analysis was thus repeated using different cut-off points at 200, 100, 60, and 30 minutes, respectively. The outcomes of the sensitivity analysis are presented in Table 6 and show the robustness of the Mann-Whitney U-Test results. Differences between males and females and day types (1 and 2) are statistically significant at all cut-off points except 30 mins with higher median values for males compared to females and for weekends & holidays compared to weekdays regardless of the cut-off point. As for interview type, schedule flexibility, and location, statistically significant differences are observed at all cut-off points with higher median values for telephone-based surveys, flexible schedules, and Zealand/Funen compared to internet-based surveys, fixed schedules, and Jutland, respectively.

**Table 5: Descriptive statistics (in minutes) and Mann-Whitney U-Test**

|  |  | Count | Mean | Std | 25% | 50% - Median | 75% | IQR |
|---|---|---|---|---|---|---|---|---|
| Gender | Male | 856 | 37.09 | 64.10 | 4.68 | 13.33 | 41.71 | 37.03 |
|  | Female | 1,072 | 31.30 | 65.85 | 3.75 | 10.17 | 25.39 | 21.64 |
|  | P-value |  |  |  |  | 0.000*** |  |  |
| Day Type 1 | Weekdays | 1,623 | 33.38 | 64.58 | 3.83 | 11.00 | 31.57 | 27.73 |
|  | Weekends | 305 | 36.49 | 68.00 | 5.72 | 13.23 | 37.15 | 31.43 |
|  |  |  |  |  |  | 0.015** |  |  |
| Day Type 2 | Weekdays | 1,502 | 32.11 | 62.83 | 3.82 | 10.78 | 29.67 | 25.85 |
|  | Weekends & Holidays | 426 | 40.09 | 72.39 | 5.61 | 13.61 | 40.63 | 35.02 |
|  |  |  |  |  |  | 0.000*** |  |  |
| Interview Type[5] | Internet | 424 | 22.85 | 48.96 | 2.85 | 7.23 | 18.33 | 15.48 |
|  | Telephone | 1,441 | 37.14 | 69.06 | 4.68 | 12.72 | 36.60 | 31.92 |
|  |  |  |  |  |  | 0.000*** |  |  |
| Schedule Flexibility | Fixed | 1,447 | 33.28 | 67.00 | 3.88 | 10.78 | 28.92 | 25.03 |
|  | Flexible | 418 | 35.37 | 59.141 | 4.55 | 14.59 | 39.65 | 35.10 |
|  |  |  |  |  |  | 0.002*** |  |  |
| Location[6] | Zealand & Funen | 1,610 | 35.42 | 66.62 | 4.52 | 12.31 | 34.63 | 30.10 |
|  | Jutland | 318 | 26.06 | 56.43 | 2.95 | 7.63 | 21.19 | 18.24 |
|  |  |  |  |  |  | 0.000*** |  |  |

** significance at the 95% level of confidence
*** significance at the 99% level of confidence

---

[5] Very few interviews are labeled as reconstructed or combined interviews instead of internet or telephone interviews and as such are not included in the analysis.
[6] Zealand, Funen, and Jutland are the three main islands in Denmark. Zealand is the most populous island in Denmark and includes the capital Copenhagen.

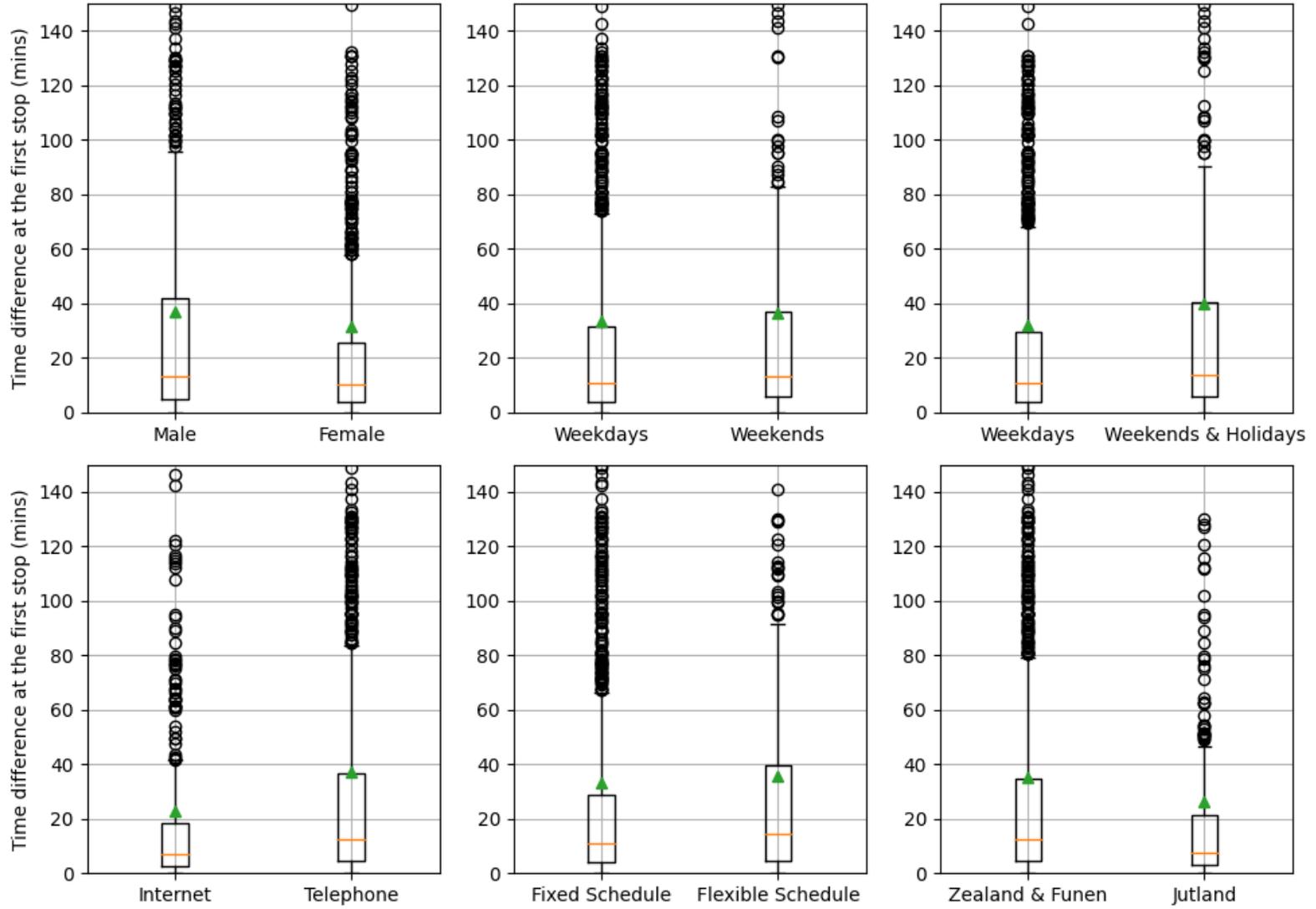

Figure 8: Box-plots of two level variables (orange line is median, green triangle is mean). The y-axis is limited to 150 minutes for visualization purposes.

Table 6: Sensitivity analysis of the Mann-Whitney U-Test w.r.t different cut-off points

| | | Cut-off point | | | | | | | | |
|---|---|---|---|---|---|---|---|---|---|---|
| | | All data | | 200 mins | | 100 mins | | 60 mins | | 30 mins | |
| | | Count | Median | Count | Median | Count | Median | Count | Median | Count | Median |
| Gender | Male | 856 | 13.33 | 804 | 12.00 | 720 | 10.63 | 622 | 8.63 | 463 | 6.47 |
| | Female | 1,072 | 10.17 | 1,006 | 9.23 | 950 | 8.44 | 880 | 7.84 | 708 | 6.29 |
| | P-value | | 0.000*** | | 0.000*** | | 0.000*** | | 0.028** | | 0.353 |
| Day Type 1 | Weekdays | 1,623 | 11.00 | 1522 | 10.01 | 1407 | 8.83 | 1276 | 7.78 | 1006 | 6.18 |
| | Weekends | 305 | 13.23 | 288 | 12.39 | 263 | 11.07 | 226 | 10.12 | 165 | 7.38 |
| | P-value | | 0.015** | | 0.002*** | | 0.004*** | | 0.026** | | 0.139 |
| Day Type 2 | Weekdays | 1,502 | 10.78 | 1,414 | 9.85 | 1,309 | 8.83 | 1,194 | 7.83 | 945 | 6.18 |
| | Weekends & Holidays | 426 | 13.61 | 396 | 12.63 | 361 | 11.07 | 308 | 9.33 | 226 | 6.96 |
| | P-value | | 0.000*** | | 0.001*** | | 0.002*** | | 0.042** | | 0.196 |
| Interview Type | Internet | 424 | 7.23 | 409 | 7.03 | 387 | 6.33 | 352 | 5.63 | 320 | 5.04 |
| | Telephone | 1,441 | 12.72 | 1,344 | 11.63 | 1,228 | 10.27 | 1,103 | 9.20 | 813 | 6.88 |
| | P-value | | 0.000*** | | 0.000*** | | 0.000*** | | 0.000*** | | 0.000*** |
| Schedule Flexibility | Fixed | 1,447 | 10.78 | 1,361 | 9.97 | 1,257 | 8.72 | 1,142 | 7.82 | 918 | 6.35 |
| | Flexible | 418 | 14.59 | 449 | 13.15 | 413 | 11.58 | 360 | 9.608 | 253 | 6.52 |
| | P-value | | 0.002*** | | 0.001*** | | 0.000*** | | 0.021** | | 0.058* |
| Location | Zealand & Funen | 1,610 | 12.31 | 1,507 | 11.08 | 1,390 | 10.13 | 1,237 | 8.72 | 940 | 6.63 |
| | Jutland | 318 | 7.63 | 303 | 7.40 | 280 | 6.70 | 265 | 6.25 | 231 | 5.17 |
| | P-value | | 0.000*** | | 0.000*** | | 0.000*** | | 0.000*** | | 0.009*** |

\* significance at the 90% level of confidence
\** significance at the 95% level of confidence
\*** significance at the 99% level of confidence

*4.2.2   Three-level+ variables*

A comparison between the three trip categories (train, bus, and mixed), different years from 2018 to 2022, and the different positions of respondents within their families was conducted using the Kruskal-Wallis H test. Results of the statistical test along with descriptive statistics are presented in Table 7. First, the Kruskal-Wallis H test shows a statistically significant difference between the three trip modes (train, bus, and mixed) at the 99% level of confidence (p = 0.000). The results also show that respondents reporting only train/metro trips are the least accurate in reporting their time of travel (12.70 minutes) while respondents reporting only bus trips are the most accurate (9.69 minutes). This discrepancy could be attributed to the higher frequencies of trains and metros, leading travelers to be more aware of bus schedules to minimize waiting times. Second, there are no statistically significant differences in the reporting error across the years from 2018 to 2022 (p-value = 0.161). Nevertheless, there is a consistent downward trend in the median value of the reporting error over the years, decreasing from 14.68 minutes in 2018 to 10.12 minutes in 2022. It is worth noting that the Rejsekort system underwent initial testing on a limited scale in 2007, but it was not until mid-2016 that it was implemented nationwide. This could potentially explain the higher median value observed in 2018, as the system was still relatively new, and people may not have been fully accustomed to it. Finally, a statistically significant difference between the different positions of respondents in their families is evident at the 99% level of confidence (p-value = 0.003). Single respondents are the least accurate with a median value of 13.33 minutes while respondents categorized as "younger in couple" are the most reliable in reporting their time of travel with a median value of 9.28 minutes. It is worth noting that around 81% of the respondents under the "younger in couple" category are females while around 78% of the "older in couple" respondents are males. This further supports the finding that "younger in couple" respondents (9.28 minutes) are more accurate than "older in couple" respondents (11.42 minutes) and is aligned with the results from the previous section (4.2.1), which indicated that females (10.17 minutes) are in general more accurate than males (13.33 minutes) (Table 5).

To ensure the reliability of the Kruskal-Wallis H test results, a sensitivity analysis was conducted with different cut-off points as done earlier in Section 4.2.1. The results presented in Table 8 confirm the findings of the Kruskal-Wallis H test. Statistically significant differences between the train, bus, and mixed travel modes are observed across all cut-off points. The bus category consistently displays the lowest median values while the train category consistently displays the highest median values. Furthermore, differences over the years are consistently not statistically significant across all cut-off points. As for the position in family, the statistical significance of the differences holds true until the 100-minute cut-off point.

*4.2.3   $1^{st}$ vs. $2^{nd}$ trip*

This section focuses on investigating the effect of reporting multiple trips on the respondents' memory. Specifically, respondents who reported 2 trips are selected for analysis. The Wilcoxon Signed Rank test was applied to test for statistical differences between reporting the first and second trip of the day. Descriptive statistics of the dependent variable, the resulting p-value of the test, and outcomes of the sensitivity analysis are presented Table 9. A statistically significant difference at the 99% level of confidence between the medians of the first and second trips (p-value = 0.002) is identified. The second trip exhibits a higher median value of 12.64 minutes compared to 10.23 minutes for the first trip suggesting that people are more accurate in reporting the start time of the first trip of the day. This could be explained by the primacy effect concept which has been widely studied in psychology and sociology. The primacy effect is a cognitive bias that refers to the tendency of people to better recall items or events that occurred or were presented at the beginning of a series or sequence (Murdock 1962). The sensitivity analysis confirms the robustness of the aforementioned finding as statistically significant differences are observed at the 99% level of confidence with higher median values for the second trip over all cut-off points.

Finally, multiple comparisons with Bonferroni adjustments show statistically significant differences at the 95% level of confidence between the categories of all variables ($p < 0.01$) except for Day Type 1 ($p = 0.015$).

**Table 7: Descriptive statistics and Kruskal-Wallis H test**

|  |  | Count | Mean | Std | 25% | 50% - Median | 75% | IQR |
|---|---|---|---|---|---|---|---|---|
| Mode | Train | 1,029 | 34.26 | 63.54 | 4.65 | 12.70 | 36.60 | 31.95 |
|  | Bus | 622 | 33.08 | 66.21 | 3.77 | 9.69 | 26.60 | 22.84 |
|  | Mixed | 277 | 34.21 | 68.63 | 3.93 | 10.50 | 24.10 | 20.17 |
|  | P-value |  |  |  |  | 0.006*** |  |  |
| Year | 2018 | 347 | 38.58 | 68.47 | 4.78 | 14.68 | 40.38 | 35.60 |
|  | 2019 | 444 | 34.91 | 67.90 | 4.38 | 12.00 | 29.27 | 24.89 |
|  | 2020 | 336 | 34.64 | 72.04 | 3.78 | 11.22 | 31.74 | 27.97 |
|  | 2021 | 362 | 28.01 | 51.63 | 4.19 | 11.28 | 24.99 | 20.80 |
|  | 2022 | 439 | 33.35 | 63.86 | 3.78 | 10.12 | 30.44 | 26.66 |
|  | P-value |  |  |  |  | 0.161 |  |  |
| Position in family | Single | 659 | 40.48 | 75.97 | 4.58 | 13.33 | 39.03 | 34.45 |
|  | Older in Couple | 391 | 31.04 | 59.21 | 4.24 | 11.42 | 33.63 | 29.39 |
|  | Younger in Couple | 461 | 26.85 | 51.78 | 3.58 | 9.28 | 24.28 | 20.70 |
|  | Child in family < 25 years | 417 | 33.85 | 64.30 | 4.12 | 11.02 | 30.63 | 26.57 |
|  | P-value |  |  |  |  | 0.003*** |  |  |

*** significance at the 99% level of confidence

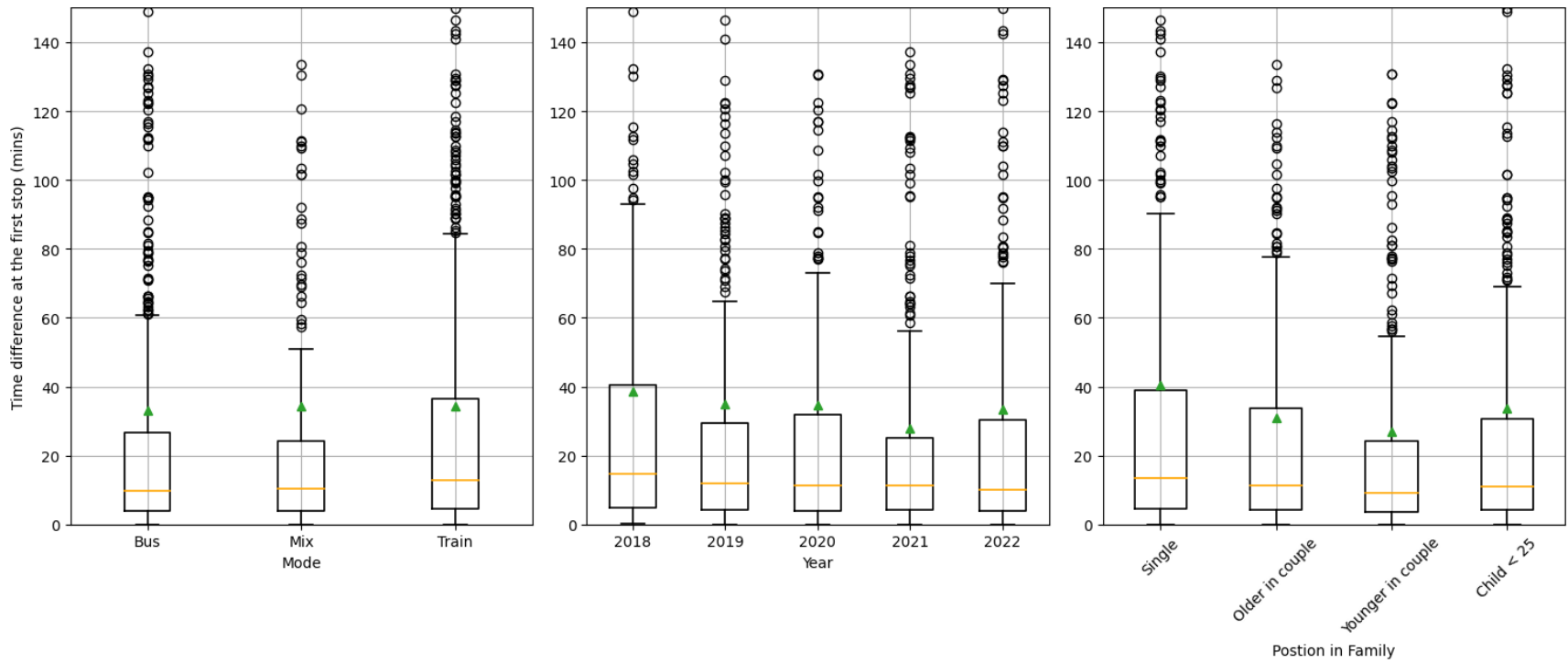

**Figure 9: Box-plots of three level+ variables (orange line is median, green triangle is mean). The y-axis is limited to 150 minutes for visualization purposes.**

**Table 8: Sensitivity Analysis of the Kruskal-Wallis H test**

| | | Cut-off point | | | | | | | | | |
|---|---|---|---|---|---|---|---|---|---|---|---|
| | | All Data | | 200 mins | | 100 mins | | 60 mins | | 30 mins | |
| | | Count | Median | Count | Median | Count | Median | Count | Median | Count | Median |
| Mode | Train | 1,029 | 12.70 | 971 | 11.72 | 908 | 10.7 | 808 | 9.66 | 567 | 6.95 |
| | Bus | 622 | 9.69 | 582 | 8.85 | 529 | 7.85 | 482 | 6.92 | 414 | 5.73 |
| | Mixed | 277 | 10.50 | 257 | 9.22 | 233 | 7.82 | 212 | 7.32 | 190 | 6.41 |
| | P-value | | 0.006*** | | 0.002*** | | 0.000*** | | 0.000*** | | 0.085* |
| Year | 2018 | 347 | 14.68 | 322 | 11.63 | 296 | 10.73 | 262 | 9.13 | 180 | 5.88 |
| | 2019 | 444 | 12.00 | 410 | 10.63 | 381 | 9.40 | 345 | 8.15 | 282 | 6.81 |
| | 2020 | 336 | 11.22 | 314 | 9.58 | 296 | 8.90 | 263 | 7.83 | 205 | 5.60 |
| | 2021 | 362 | 11.28 | 347 | 11.00 | 321 | 10.05 | 293 | 8.85 | 232 | 6.97 |
| | 2022 | 439 | 10.12 | 417 | 9.68 | 376 | 8.29 | 339 | 7.43 | 272 | 6.31 |
| | P-value | | 0.161 | | 0.285 | | 0.11 | | 0.15 | | 0.35 |
| Position in Family | Single | 659 | 13.33 | 602 | 12.00 | 550 | 10.41 | 500 | 9.67 | 373 | 6.95 |
| | Older in Couple | 391 | 11.42 | 377 | 10.85 | 353 | 9.68 | 303 | 8.15 | 232 | 6.29 |
| | Younger in Couple | 461 | 9.28 | 439 | 8.83 | 409 | 7.82 | 387 | 6.98 | 317 | 6.00 |
| | Child < 25 years | 417 | 11.02 | 392 | 10.35 | 358 | 8.92 | 312 | 7.51 | 249 | 6.45 |
| | P-value | | 0.003*** | | 0.029** | | 0.025** | | 0.112 | | 0.658 |

\* significance at the 90% level of confidence
\*\* significance at the 95% level of confidence
\*\*\* significance at the 99% level of confidence

**Table 9: Descriptive statistics and Wilcoxon Signed Rank test**

|  | Count | Mean | Std | 25% | 50% - Median | 75% | IQR |
|---|---|---|---|---|---|---|---|
| **All data** | | | | | | | |
| 1st Trip | 898 | 32.11 | 64.42 | 3.78 | 10.23 | 30.99 | 27.22 |
| 2nd Trip | 898 | 35.88 | 66.03 | 4.57 | 12.64 | 34.31 | 29.74 |
| P-value | | | | | 0.002*** | | |
| **Cut-off point – 200 mins** | | | | | | | |
| 1st Trip | 842 | | | | 9.38 | | |
| 2nd Trip | 842 | | | | 11.60 | | |
| P-value | | | | | 0.003*** | | |
| **Cut-off point – 100 mins** | | | | | | | |
| 1st Trip | 781 | | | | 8.37 | | |
| 2nd Trip | 781 | | | | 10.28 | | |
| P-value | | | | | 0.007*** | | |
| **Cut-off point – 60 mins** | | | | | | | |
| 1st Trip | 703 | | | | 7.08 | | |
| 2nd Trip | 703 | | | | 9.23 | | |
| P-value | | | | | 0.002*** | | |
| **Cut-off point – 30 mins** | | | | | | | |
| 1st Trip | 551 | | | | 5.63 | | |
| 2nd Trip | 551 | | | | 7.40 | | |
| P-value | | | | | 0.001*** | | |

*** significance at the 99% level of confidence

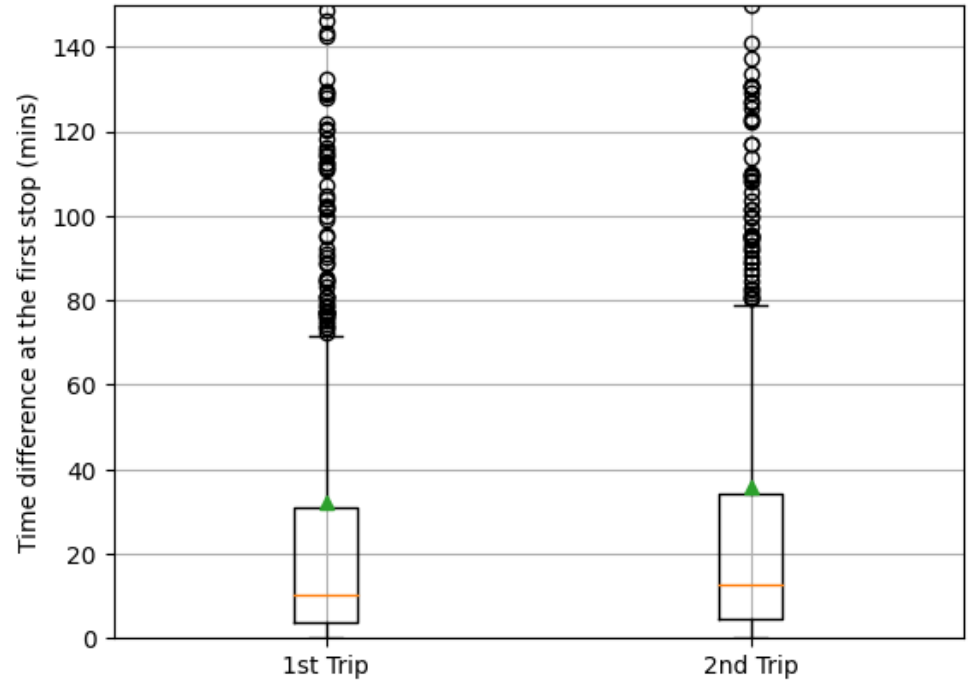

**Figure 10: Box-plot of 1st vs. 2nd trip (orange line is median, green triangle is mean). The y-axis is limited to 150 minutes for visualization purposes.**

## 4.3 Early or Late Reporting?

In addition to examining the absolute time difference, we also analyze the instances of early- and late-reporting of time differences at both the first and last stops of a trip. Early-reporting is indicated by a negative ΔT (the difference between Tap-in/Tap-out times from Rejsekort and reported arrival times in TU) at either the first or last stop while late-reporting is indicated by a positive ΔT. Approximately 47% of the reported trips show a negative ΔT at the first stop, while 53% display a positive ΔT. These proportions are consistent at the last stop as well. Moreover, Figure 11 shows a symmetrical normal distribution, centered around 0, of the time difference, $ΔT_{n,First\_St}$, at the first stop/stations. Figure 12 shows that the differences between the reported times at the first and last stops are positively correlated in a consistent linear pattern. This pattern indicates that if there is a significant time difference at the start of the trip, a similar time difference is observed at the end of the trip. Moreover, the heatmap matrix (Figure 13) further explores the distribution of trips based on the sign of the time differences (ΔT) at the first and last stops. It indicates that the majority of trips (85.73%) show consistent reporting patterns, either late-reporting (40.15%) or early-reporting (45.58%) at both stops. A smaller proportion of trips (14.27%) exhibit inconsistent reporting, with late-reporting at one stop and early-reporting at the other. This consistency reinforces the linear pattern observed in Figure 12, suggesting that discrepancies in reported times are systematic and consistent across the entire trip.

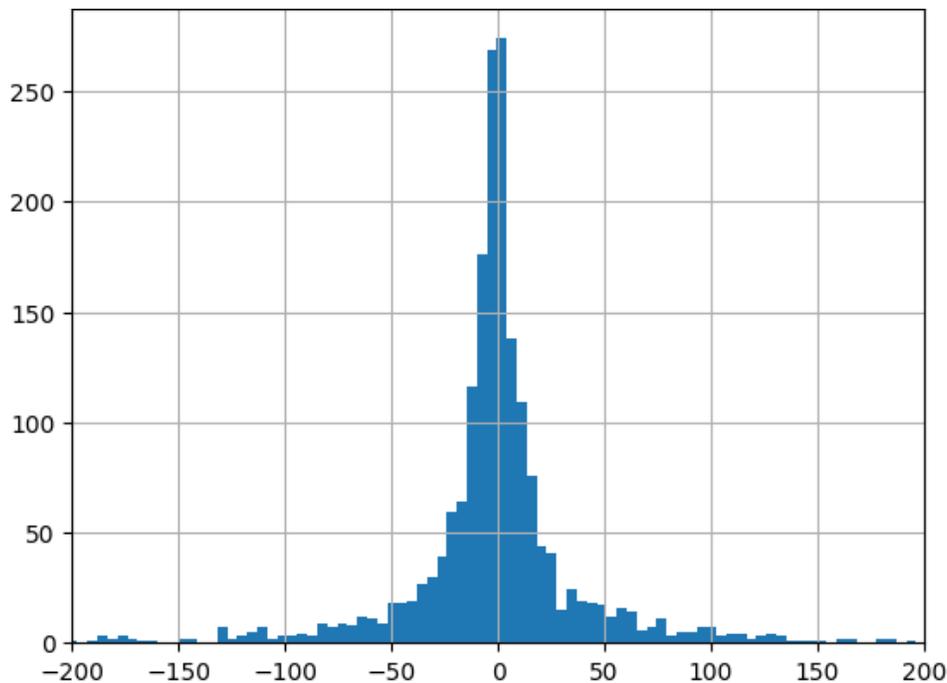

**Figure 11: Distribution of time differences at first stop/station $ΔT_{n,First\_St}$ (in minutes). The x-axis is limited to (-200, 200) minutes for visualization purposes.**

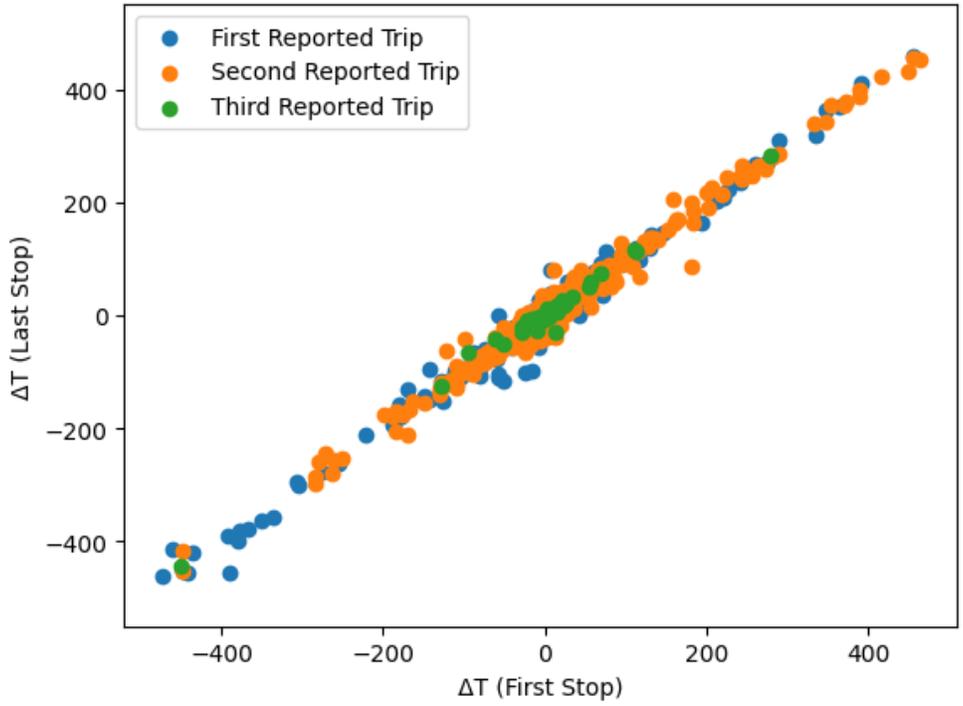

Figure 12: Correlation of time differences at first and last stop (in minutes)

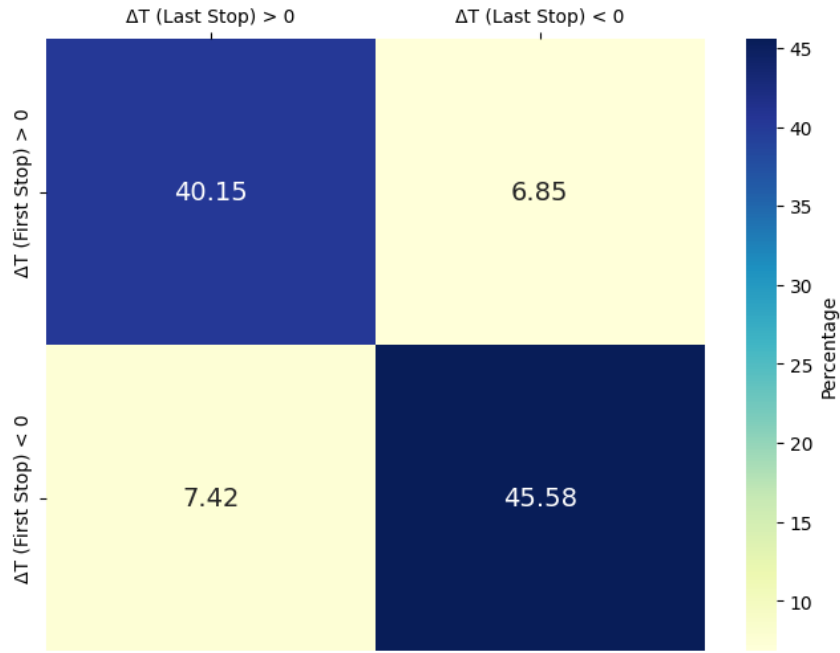

**Figure 13: Heatmap of over- and under- reporting at first and last stops**

## 4.4 Case of Second Matched Card

Approximately 16% of individuals who reported two or three trips in the travel survey had only one unique match in the smart card data. About 30% had at least two matches, with the average time difference between the first and second matched card being less than 5 minutes. Around 19% had at least two matches with a time difference between the first and second match exceeding one hour (Table 10). A further sensitivity analysis was performed to verify the robustness of the conclusions from section 4.2. For individuals with at least two matched cards and an average time difference of less than 5 minutes, the second matched card was chosen, and the statistical analysis was repeated on the entire matched sample. The conclusions remained consistent, showing similar statistical differences across all categories, with slightly higher median values (Tables 11 and 12). By considering the first matched card, the reporting error measures in sections 4.1 and 4.2 can be regarded as a lower bound of the true error. This conservative approach would result in an underestimation of the error in case of card mismatches, rather than an overestimation.

**Table 10: Second matched card**

|  | Average Time difference between first and second matched card (minutes) | Count | % |
|---|---|---|---|
| Individuals with 1 matched card |  | 147 | 15.61% |
| Individuals with at least 2 matched smart cards | ]0-5[ | 281 | 29.83% |
|  | [5-30[ | 252 | 26.75% |
|  | [30-60[ | 84 | 8.92% |
|  | 60+ | 178 | 18.89% |

**Table 11: Descriptive statistics (in minutes) and Mann-Whitney U-Test – case of second matched card**

|  |  | Count | Mean | Std | 25% | 50% - Median | 75% | IQR |
|---|---|---|---|---|---|---|---|---|
| Gender | Male | 856 | 37.48 | 64.39 | 5.18 | 13.68 | 41.99 | 36.81 |
|  | Female | 1,072 | 31.70 | 65.74 | 4.10 | 10.74 | 25.37 | 22.28 |
|  | P-value |  |  |  |  | 0.000*** |  |  |
| Day Type 1 | Weekdays | 1,623 | 33.80 | 64.67 | 4.23 | 11.75 | 32.32 | 28.08 |
|  | Weekends | 305 | 36.79 | 67.99 | 5.82 | 13.40 | 37.15 | 31.33 |
|  |  |  |  |  |  | 0.030** |  |  |
| Day Type 2 | Weekdays | 1,502 | 32.57 | 62.95 | 4.17 | 11.50 | 30.60 | 26.43 |
|  | Weekends & Holidays | 426 | 40.26 | 72.33 | 5.83 | 13.63 | 41.38 | 35.56 |
|  |  |  |  |  |  | 0.002*** |  |  |
| Interview Type[7] | Internet | 424 | 22.94 | 48.97 | 2.83 | 7.08 | 18.38 | 15.55 |
|  | Telephone | 1,441 | 37.60 | 69.14 | 5.22 | 13.50 | 37.25 | 32.03 |
|  |  |  |  |  |  | 0.000*** |  |  |
| Schedule Flexibility | Fixed | 1,447 | 33.76 | 67.00 | 4.16 | 11.38 | 29.88 | 25.72 |
|  | Flexible | 418 | 38.58 | 64.03 | 4.55 | 15.97 | 44.69 | 40.14 |
|  |  |  |  |  |  | 0.008*** |  |  |
| Location[8] | Zealand & Funen | 1,610 | 35.68 | 66.65 | 4.59 | 12.73 | 35.22 | 30.63 |
|  | Jutland | 318 | 27.14 | 56.79 | 3.37 | 8.84 | 22.96 | 19.59 |
|  |  |  |  |  |  | 0.000*** |  |  |

** significance at the 95% level of confidence
*** significance at the 99% level of confidence

---

[7] Very few interviews are labeled as reconstructed or combined interviews instead of internet or telephone interviews and as such are not included in the analysis.
[8] Zealand, Funen, and Jutland are the three main islands in Denmark. Zealand is the most populous island in Denmark and includes the capital Copenhagen.

Table 12: Descriptive statistics and Kruskal-Wallis H test - case of second matched card

|  |  | Count | Mean | Std | 25% | 50% - Median | 75% | IQR |
|---|---|---|---|---|---|---|---|---|
| Mode | Train | 1,029 | 34,50 | 63,62 | 5,00 | 13,17 | 36,62 | 31,62 |
|  | Bus | 622 | 33,85 | 66,31 | 4,00 | 10,85 | 28,37 | 24,36 |
|  | Mixed | 277 | 34,35 | 68,60 | 4,00 | 10,80 | 24,95 | 20,95 |
|  | P-value |  |  |  |  | 0.012** |  |  |
| Year | 2018 | 347 | 38,69 | 68,44 | 5,03 | 15,28 | 40,08 | 35,06 |
|  | 2019 | 444 | 35,57 | 68,16 | 4,68 | 12,64 | 32,25 | 27,57 |
|  | 2020 | 336 | 35,26 | 71,89 | 4,30 | 12,30 | 32,20 | 27,91 |
|  | 2021 | 362 | 28,05 | 52,03 | 4,15 | 10,78 | 25,92 | 21,77 |
|  | 2022 | 439 | 33,84 | 63,77 | 3,77 | 10,72 | 33,23 | 29,46 |
|  | P-value |  |  |  |  | 0.152 |  |  |
| Position in family | Single | 659 | 40,82 | 76,03 | 4,80 | 13,58 | 39,12 | 34,32 |
|  | Older in Couple | 391 | 31,35 | 59,20 | 4,53 | 12,10 | 33,78 | 29,26 |
|  | Younger in Couple | 461 | 27,05 | 51,56 | 4,13 | 9,73 | 24,27 | 20,13 |
|  | Child in family < 25 years | 417 | 34,64 | 64,69 | 4,28 | 11,63 | 32,18 | 27,90 |
|  | P-value |  |  |  |  | 0.005*** |  |  |

** significance at the 95% level of confidence
*** significance at the 99% level of confidence

# 5   CONCLUSION

This paper quantified the time reporting error of public transit users in a nationwide household travel survey by matching five years of data from two sources, the Danish National Travel Survey (TU) and the Danish Smart Card system (Rejsekort). The time reporting error corresponds to the absolute time difference between the reported time in the TU data and the tap-in time in the Rejsekort data at the first stop of a trip. Around 70% of TU respondents who reported 2 or 3 public transport trips were successfully matched with travel cards from the Rejsekort data solely based on respondents' declared travel behavior and tap-in/tap-out transactions. The reporting error had a median of 11.34 minutes with an Interquartile Range of 28.14 minutes. In addition, the paper investigated the relationships between the survey respondents' reporting error and their socio-economic and demographic characteristics using non-parametric statistical tests. The results showed that males are in general less accurate than females in reporting their times of travel. Respondents with a flexible schedule are also less accurate than those with a fixed schedule. Moreover, trips reported during weekends and holidays, via telephones, or from Zealand/Funen displayed lower accuracies compared to trips reported during weekdays, via the internet, or from Jutland, respectively. Furthermore, the results showed that respondents are more likely to accurately remember their times of travel by bus as opposed to train or metro. The difference between the median reporting errors across the categories of each variable varies between 5.48 and 2.23 minutes. Ranking the variables by the size of the difference puts interview type at the top of the list with 5.48 minutes followed by location with 4.68 minutes and position in family with 4.05 minutes. Subsequently, schedule flexibility, gender, and mode follow with a difference of 3.81, 3.16, and 3.01 minutes, respectively. Finally, Day Type 1 (weekdays vs. weekends), $1^{st}$ vs. $2^{nd}$ trip, and Day Type 2 (weekdays vs. weekends & holidays) complete the list with differences of 2.83, 2.42, and 2.23 minutes, respectively. The findings highlight the importance of considering individual-level comparison between travel surveys and smart card data, as such comparison offers a better understanding of reporting errors in travel surveys and their connections to different socio-economic and demographic characteristics.

It is hoped that quantifying and understanding reporting errors in travel surveys could help policymakers and researchers in two significant ways: improving data collection and correcting data post-collection. By identifying categories associated with higher reporting errors, data collectors can implement additional probing or data checks with respondents during the survey process, helping to obtain more accurate information from groups prone to reporting inaccuracies. After collecting the data, it is possible to correct for reporting errors by replacing components likely to contain errors with matched data from smart cards. This method allows the continued use of survey data while enhancing its accuracy by integrating reliable smart card data for components prone to reporting errors. Doing so, would enhance the accuracy of activity-based models. For instance, significant reporting errors would affect the accuracy of time-of-day choice models, especially if departure time choice alternatives are modeled at the level of 15 or 30 minutes. Errors in departure times can accumulate, leading to deviations in the estimation of peak travel times. It can also lead to the underestimation and overestimation of travel demand during certain periods.

Overall, quantifying and understanding reporting errors in travel surveys offers a pathway to enhance the accuracy and reliability of such data. These insights can refine data collection techniques and improve data quality by considering the socio-economic and demographic characteristics that have been investigated. This study is not devoid of limitations. Firstly, TU respondents who only reported one trip per day were excluded from the analysis due to the high probability of identifying multiple matches in the Rejekort data. Future work could focus on developing effective heuristics to address this matching challenge. Secondly, a more in-depth investigation of the unmatched observations should be conducted. Understanding the reasons behind unsuccessful matches could offer valuable information about any potential biases in the data. It would also help in improving data collection techniques. Finally, the assumption of travelers tapping in upon arrival at train/metro stations might not hold true in all cases. The variability in the timing of tap-ins between arrival and boarding time should be further investigated to evaluate its impact on the robustness of the findings.

## ACKNOWLEDGEMENTS

The research leading to these results has received funding from the Horizon Europe Framework Programme under the Marie Skłodowska-Curie Postdoctoral Fellowship MSCA-2021-PF-01 project No 101063801.

## AUTHOR CONTRIBUTIONS

The authors confirm contribution to the paper as follows: study conception and design: Georges Sfeir, Filipe Rodrigues; data collection and software: Georges Sfeir; analysis and interpretation of results: Georges Sfeir, Filipe Rodrigues, Francisco Pereira, Maya Abou Zeid; draft manuscript preparation: Georges Sfeir, Filipe Rodrigues, Francisco Pereira, Maya Abou Zeid. All authors reviewed the results and approved the final version of the manuscript.

# APPENDIX A

The days on which respondents reported PT trips in the TU survey were selected for the calculation of mean and median values of number of trips per day and for the below plots.

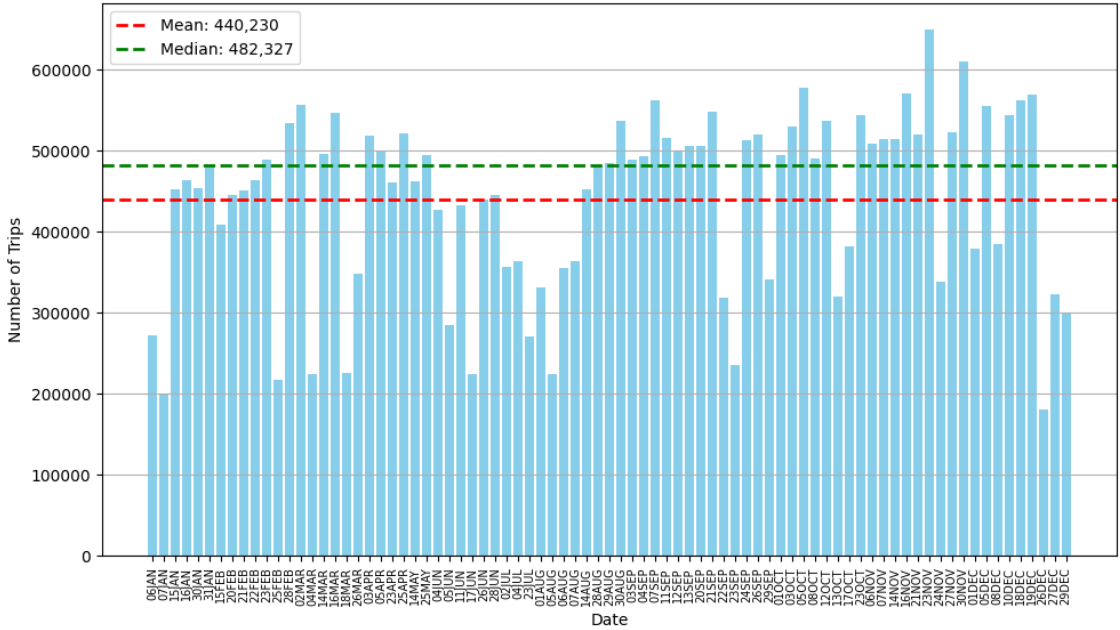

**Figure 14: Number of trips per day (2018)**

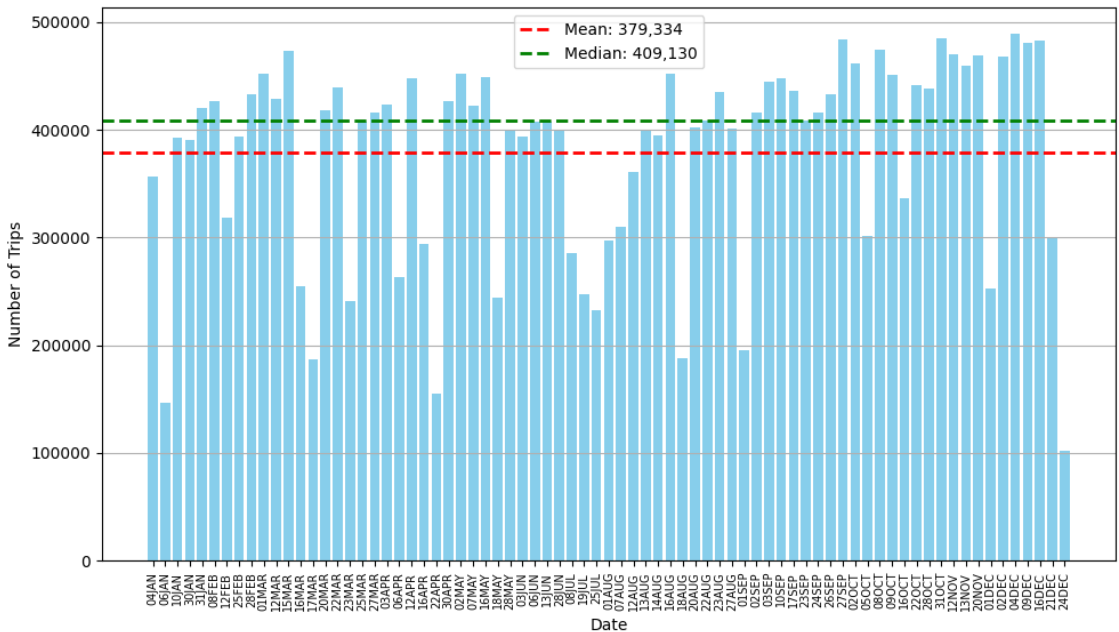

**Figure 15: Number of trips per day (2019)**

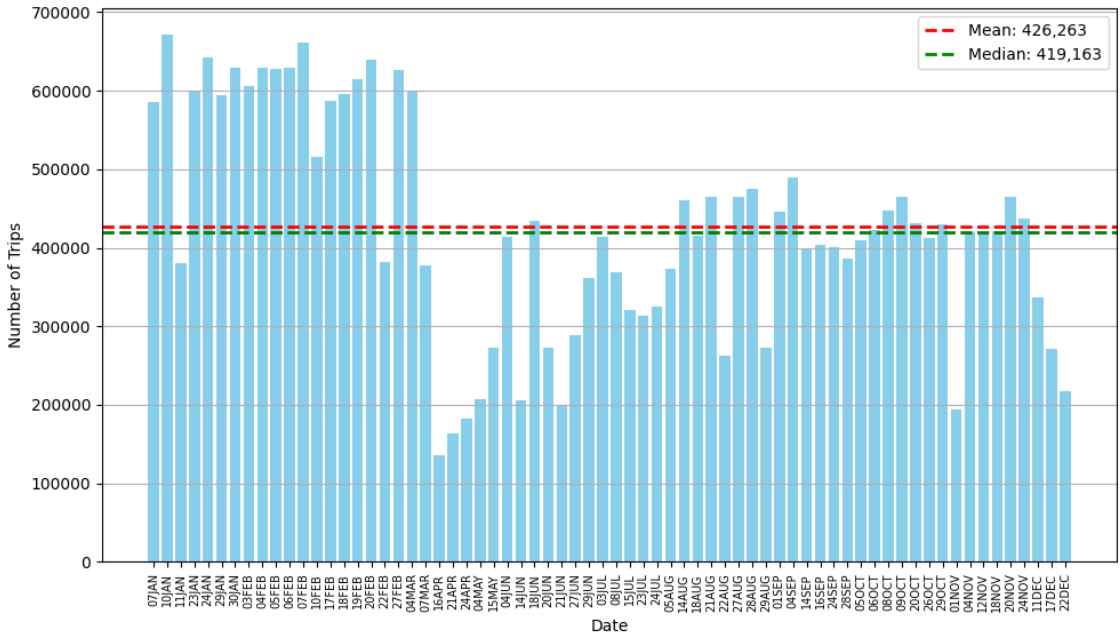

**Figure 16: Number of trips per day (2020)**

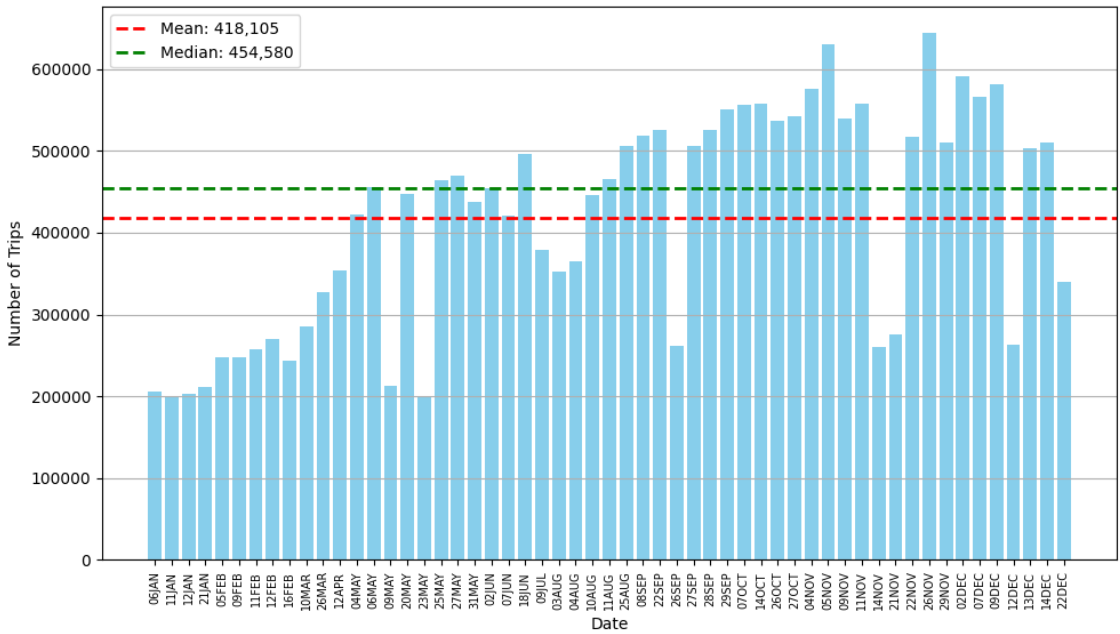

**Figure 17: Number of trips per day (2021)**

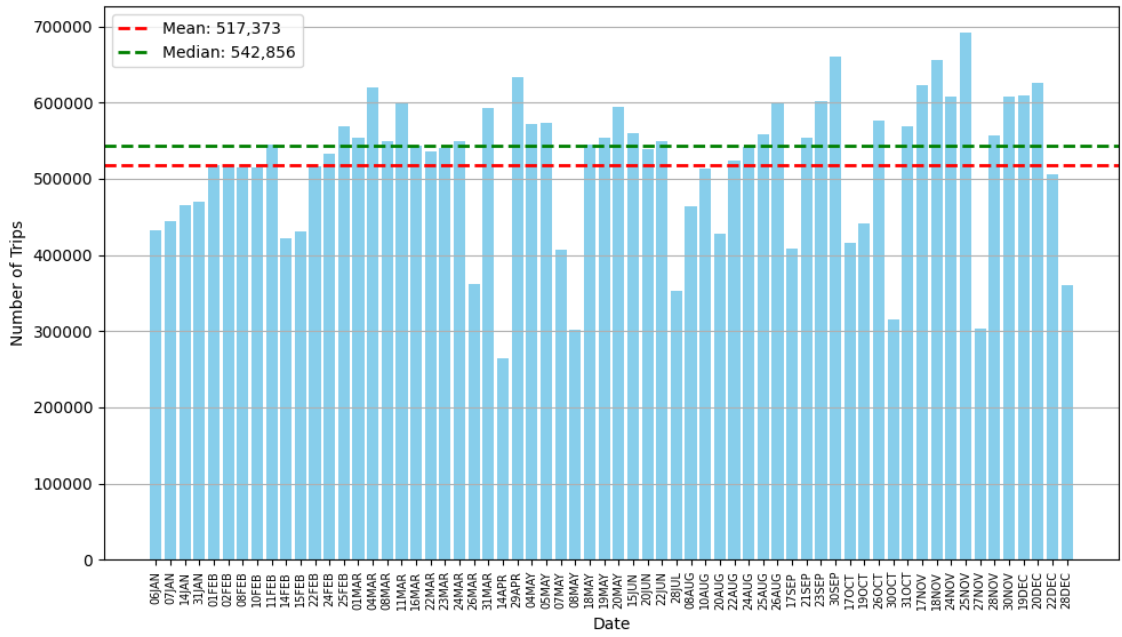

**Figure 18: Number of trips per day (2022)**